\documentclass[letterpaper]{article}

\usepackage[T1]{fontenc}
\usepackage[latin9]{inputenc}
\usepackage[english]{babel}
\usepackage{amsmath}
\usepackage{graphicx}
\usepackage{multirow}
\usepackage{xcolor}
\usepackage[numbers,sort&compress]{natbib}
\usepackage{hyperref}
\usepackage{vmargin}
\usepackage[pass,papersize={8.5in,11in}]{geometry}
\usepackage{authblk}

\pdfpagewidth=\paperwidth
\pdfpageheight=\paperheight
\setpapersize{USletter}
\setmarginsrb{1in}{1in}{1in}{1in}{0.25in}{0.25in}{0.25in}{0.25in}

\title{In-situ Characterization of the Thermal State of Resonant Optical Interferometers via 
	Tracking of their Higher-order Mode Resonances}

\author[1)]{Chris L. Mueller}
\author[1)]{Paul Fulda}
\author[2)]{Rana X. Adhikari}
\author[2)]{Koji Arai}
\author[3)]{Aidan F. Brooks}
\author[2)]{Rijuparna Chakraborty}
\author[4)]{Valery V. Frolov}
\author[5)]{Peter Fritschel}
\author[6)]{Eleanor J. King}
\author[1)]{David B. Tanner}
\author[3)]{Hiroaki Yamamoto}
\author[1)]{Guido Mueller}

\affil[1)]{Department of Physics, University of Florida, Gainesville, FL 32611, USA}
\affil[2)]{Department of Physics, California Institute of Technology, Pasadena, CA 91125, USA}
\affil[3)]{LIGO, California Institute of Technology, Pasadena, CA 91125, USA}
\affil[4)]{LIGO Livingston Observatory, Livingston, LA 70754, USA}
\affil[5)]{LIGO, Massachusetts Institute of Technology, Cambridge, MA 02139, USA}
\affil[6)]{University of Adelaide, Adelaide, SA 5005, Australia}

\begin{document}

% Make Title
\maketitle

% \tcr{Possible Journals: PRD, CQG, Applied Optics, JOSA B}

%==================================================================================================
% Abstract
%==================================================================================================
\abstract{
	Thermal lensing in resonant optical interferometers such as those used for gravitational 
	wave detection is a concern due to the 
	negative impact on control signals and instrument sensitivity. 
	In this paper we describe a method for monitoring the thermal state of such interferometers by 
	probing the higher-order spatial mode resonances of the cavities within them. 
	We demonstrate the use of this technique to measure changes in the Advanced LIGO input mode 
	cleaner cavity geometry as a function of input power, and subsequently infer the optical 
	absorption at the mirror surfaces at the level of $1\ \text{ppm}$ per mirror. 
	We also demonstrate the generation of a useful error signal for thermal state of the Advanced LIGO power 
	recycling cavity by continuously tracking the first order spatial mode resonance frequency. 
	Such an error signal could be used as an input to thermal compensation systems to maintain the 
	interferometer cavity geometries in the presence of transients in circulating light power 
	levels, thereby maintaining optimal sensitivity and maximizing the duty-cycle of the detectors.
}

%==================================================================================================
% Introduction
%==================================================================================================
\section{Introduction}
\label{sec:Introduction}

The second generation of kilometer-scale ground based laser interferometric gravitational wave 
detectors is currently under development, with aims of reaching unprecedented strain sensitivities 
and making the first direct detection of gravitational waves. 
Advanced LIGO (aLIGO) in the US consists of two 4\,km dual-recycled Fabry-Perot Michelson 
interferometers situated in Hanford, Washington and Livingston, 
Louisiana~\cite{harry_advanced_2010}. 
aLIGO incorporates many upgrades over the initial LIGO project\cite{abbott_ligo_2009}, including improved seismic 
isolation and suspensions for the core optics, the addition of a signal recycling mirror, stable 
recycling cavities and an increase in the circulating light power. 
The increased circulating light power in particular places great importance on the ability to 
characterize the thermal state of the interferometer, because of the effects of thermal lensing 
associated with absorption in the core optics~\cite{Strain_thermal_1994}.  

Optical cavities, such as those which make up the aLIGO interferometers, have a defined set of 
resonances which may be characterized by both their longitudinal spacing and transverse structure. 
For any given longitudinal mode, an infinite set of transverse modes  
describes the spatial properties of the beam in a plane perpendicular to the propagation axis. 
In the case of an optical cavity with spherically curved mirrors, the higher-order transverse modes 
(HOMs) are well described by the Hermite-Gauss functions~\cite{bayer-helms_coupling_1984}. 
Measuring the eigenfrequencies of these HOMs provides a powerful method of characterizing 
an optical cavity. 

Measurement of the spacing between the fundamental resonances has been used to measure the length 
of an optical resonator to a precision of 
$\sim5\,\text{ppm}$~\cite{stochino_technique_2012,mueller_techniques_2014}.  
Precise measurement of the width of the fundamental resonance, the so-called cavity pole, provides 
a measurement of the losses in a high finesse resonator with an accuracy of 
$2\,\text{ppm}$~\cite{uehara_accurate_1995}.  
Measurement of the frequency spacing between the fundamental mode and the first excited modes 
can be used to measure the cavity Gouy phase to a level of 
$\sim2\cdot10^{-2}$~\cite{mueller_techniques_2014} from which the radii of curvature of the 
mirrors can be inferred with additional information~\cite{stochino_technique_2012}.  

A number of other complementary techniques have been developed to characterize resonant 
interferometers.  
The dynamic response of optical cavities to laser frequency variations has been used to measure 
the length of the LIGO arms to a precision of $80\ \mu\text{m}$~\cite{rakhmanov_characterization_2004}.  
Their dynamic response to amplitude variations has been used to measure the round-trip loss at the 
10 ppm level~\cite{isogai_loss_2013}.

In this paper we report on the methods and results of two experiments which trace the thermal 
state of an optical cavity by monitoring the spacing between the fundamental mode and the first 
order spatial mode.  
In particular, this frequency spacing is used to infer the amount of total absorption in the 
aLIGO input mode cleaner (IMC) as well as to generate an error signal for thermal 
compensation of the aLIGO power recycling cavity (PRC), both of which are located at the LIGO 
Livingston Observatory (LLO).  

The paper is organized as follows; 
Section \ref{sec:measurement_principle} gives a general description of the technique used to 
measure the resonant frequencies of an optical cavity.  
Section \ref{sec:calcs} explains how one calculates the eigenspectrum of an ideal resonator, and 
Section \ref{sec:thermal} shows how tracking of the first order resonance can be used to sense the 
thermal state of a resonator.    
Section \ref{sec:imc} describes how this technique was applied to measure the absorption of the 
aLIGO IMC.  
Section \ref{sec:prc} describes how the technique was used to generate an error signal of 
the thermal state of the aLIGO PRC.  
Finally, Section \ref{sec:conclusion} closes with some conclusions and an outlook of how this 
technique might be applied in the future.

%==================================================================================================
% Measurement Principle
%==================================================================================================
\section{Measurement principle}
\label{sec:measurement_principle}

In order to measure the eigenfrequencies of an optical cavity it is first necessary to control 
the resonance condition of the cavity for a fixed `carrier' frequency. 
Control is typically accomplished with some variant of the Pound-Drever-Hall (PDH)  
technique~\cite{drever_laser_1983,black_introduction_2001}, whereby an error signal for the carrier 
frequency resonance condition is generated from the beat between fixed frequency RF phase modulation 
sidebands and the carrier light in reflection from the cavity. This error signal is applied as 
feedback to stabilize the relative fluctuations between the frequency of the laser and the length 
of the cavity.  

An additional sideband (or pair of sidebands) with a tunable frequency offset to the carrier 
frequency may then be used to probe the eigenspectrum of the cavity. 
This can be accomplished by adding a pair of sidebands to the carrier beam with phase or amplitude 
modulation, or by injecting an auxiliary laser beam which is phase locked to the carrier beam.  
Using modulation has the advantage that the sideband fields are already aligned to the carrier 
beam, while using an auxiliary laser has the advantage of adding only a single sideband to the 
carrier beam, thus simplifying the analysis. 
Henceforth we do not consider the fixed frequency sidebands used for PDH control in this 
discussion; the term sidebands will be used only to describe the tunable frequency sidebands used 
for probing the cavity eigenspectra. 

For the case of a tunable single sideband we may write the beam incident upon the cavity as
\begin{equation}
	E_0=E_ce^{-i\omega_0t}\sum_{nm}b_{nm}U_{nm}+E_se^{-i(\omega_0+\Omega)t}\sum_{nm}c_{nm}U_{nm},
\end{equation}
where $E_c$ and $\omega_0$ are the field amplitude and frequency of the carrier, $E_s$ is the 
field amplitude of the sideband, $\Omega$ is the offset frequency between the carrier and the 
sideband (the modulation frequency), $U_{nm}$ are the functions describing the spatial structure of the HOMs of the 
cavity, and $|b_{nm}|^2$ ($|c_{nm}|^2$) quantifies how much carrier (sideband) power is in each 
of the HOMs. 

When the offset frequency of the sideband is near a resonance for particular HOM, say the 
$n^\prime m^\prime$ mode, the beam transmitted through the cavity will be
\begin{equation}
	E(t)=t(\omega_0)E_ce^{-i\omega_0t}U_{00}+t(\omega_0+\Omega)E_se^{-i(\omega_0+\Omega)t}c_{n'm'}U_{n'm'},
\end{equation}
where we have assumed for simplicity of exposition that the cavity is non-degenerate with a high 
finesse such that all non-resonant modes are completely reflected. 
Here $t(\omega_0)$ is the transmittance of the cavity to the $00$ mode of the carrier which is assumed 
to be real, $t(\omega_0+\Omega)$ is the transmittance of the cavity at frequency $\omega_0+\Omega$ associated with 
which is near the resonance of the $n'm'$ mode.

Calculating the power in this beam gives
\begin{equation}
	P=t(\omega_0)^2E_c^2|U_{00}|^2+|t(\omega_0+\Omega)|^2E_s^2c_{n'm'}^2|U_{n'm'}|^2
		+2t(\omega_0)^2E_cE_sc_{n'm'}\Re\left[U_{n'm'}U_{00}^*t(\omega_0+\Omega)\right].
\end{equation}
Finally, integrating spatially over the incomplete cross section of the beam $\partial$ and 
extracting the terms which will show up after demodulation at the sideband offset frequency 
$\Omega$ gives
\begin{equation}
	P_\Omega=2t(\omega_0)^2E_cE_sc_{n'm'}\int_\partial U_{n'm'}U_{00}^*\left\{\Re[t(\omega_0+\Omega)]
		\cos(\Omega t)+\Im[t(\omega_0+\Omega)]\sin(\Omega t)\right\}\ dA,
		\label{eq:experimental_description_power_at_omega}
\end{equation}
where we have assumed that the Gaussian beam parameters are the same for the sideband and the 
carrier. 

The transmittance coefficient for a two mirror spherical cavity is given by 
\begin{equation}
	t(\omega)=\frac{t_1t_2e^{-i\frac{L}{c}(\omega-\Omega_0)}}
		{1-r_1r_2e^{-i2\frac{L}{c}(\omega-\Omega_0)}},
\end{equation}
where $r_i$ and $t_i$ are the reflectance and transmittance of the mirrors, $L$ is the one 
way length of the cavity, and $\Omega_0$ is the resonant frequency of the particular HOM.  
This function undergoes a rapid phase transition as $\Omega$ is swept through the resonance, 
allowing for a precise determination of the resonance frequency.  
More complicated and realistic optical resonators, including the aLIGO IMC have similar transmittance coefficients near a 
resonance.  

From this result we can see that the two quadratures of the demodulated signal map out both 
quadratures of the complex transmittance coefficient of the optical cavity at each HOM resonance. 
The phase flip in the complex transmittance coefficient across the resonance of a given 
higher-order spatial mode gives an ideal signal for precisely determining the eigenfrequency of 
that mode. 

The calculation can be generalized to show that the measurement principle still works with 
multiple sidebands, as in the case where a modulator is used rather than an auxiliary laser.  
In addition, a similar calculation shows that the measurement can also be made in reflection of an 
optical cavity with the only major difference being that the complex reflectance coefficient of 
the cavity is measured rather than the transmittance coefficient. 

Notice that the signal in Eqn.~\eqref{eq:experimental_description_power_at_omega} is 
proportional both to the amount of HOM power injected, captured in the coefficient 
$c_{n'm'}$, and the amount of HOM power detected, captured in the integral. 
The creation of higher-order modes in the sideband field is most easily achieved by partially occulting the  
input beam, thereby scattering the mostly Gaussian (HG$_{00}$) beam into a 
range of higher-order modes. 
This method has the advantage of creating many HOMs so that the various resonances 
of the optical cavity can be probed without any significant changes to the experimental setup.  

It is also important that the measurement device is sensitive to the beat note between the carrier 
beam, which is in the HG$_{00}$ mode, and the components of the sideband beam which are in 
higher-order spatial modes HG$_{nm}$. 
Integrating over the full spatial profile of the beam by focusing it onto a photodiode would 
cause the experiment to be insensitive to this beat note, due to the orthogonality of 
the HG modes over an infinite transverse plane. 
The detection method employed in our experiments was to simply occult a portion of the beam 
before focusing it onto the photodetector, thus breaking the symmetry of the infinite transverse plane  
and providing sensitivity to a wide range of HOMs.

%==================================================================================================
% Eignespectrum of an Ideal Resonator
%==================================================================================================
\section{The eigenspectrum of an ideal resonator}
\label{sec:calcs}

Monochromatic radiation picks up an additional phase shift relative to a simple plane wave, known 
as the Gouy phase~\cite{siegman_lasers_1986}.  
The amount of Gouy phase accumulated upon propagation is dependent on the spatial structure of the beam, 
but can be expressed rather simply when the beam is decomposed in the HG basis. 
In this basis, the Gouy phase accumulated by each HOM is related to that accumulated by the 
fundamental Gaussian ($n=m=0$) mode simply by
\begin{equation}
	\Psi_{nm}=(n+m+1)\ \Psi_{00},
\end{equation}
where $n$ and $m$ specify the mode indices in the HG basis. 

The geometry of a spherical optical cavity defines the modal basis of the beam that resonates 
within it. 
The accumulated round trip Gouy phase is therefore also defined by the cavity geometry. 
For a simple two mirror cavity the fundamental round-trip Gouy phase is given simply by
\begin{equation}
	\Psi_{00}=2\cos^{-1}\left(\sqrt{g_1\ g_2}\right)
	\label{eq:calcs_gouy_fp}
\end{equation}
where $g_i$ are the standard cavity stability parameters given by
\begin{equation}
	g_i=1-\frac{L}{R_i},
	\label{eq:calcs_g_factor}
\end{equation}
where $L$ is one half of the round-trip length of the cavity and $R_i$ are the radii of curvature 
of the two mirrors. 

The Gouy phase for a more complex cavity consisting of more than two mirrors is most easily 
calculated with ray matrix methods~\cite{siegman_lasers_1986,shaomin_matrix_1985}.  
The fundamental round-trip Gouy phase can be obtained from the  
expression~\cite{arai_accumulated_2013,erden_accumulated_1997}
\begin{equation}
	\Psi_{00}=2\cos^{-1}\left(\frac{B}{|B|}\sqrt{\frac{A+D+2}{4}}\right),
	\label{eq:calcs_gouy_complex}
\end{equation}
where $A$, $B$, and $D$ are three elements of the round-trip ABCD matrix.  

The length of the cavity determines the frequency spacing between the longitudinal modes  
(successive resonances of any given transverse mode). 
This parameter, known as the free spectral range (FSR), is given for a cavity of round-trip length $L$ by
\begin{equation}
	f_0=\frac{c}{L}.
\end{equation} 
In the experiments which follow the fundamental mode of the carrier will be held on resonance by a 
control system while the frequency of the sideband is shifted relative to this frequency.  
The frequency location of the resonances of the HOMs of an optical cavity 
are therefore given by
\begin{equation}
	f_{nm}=\frac{c}{2\pi L}\text{mod}_{2\pi}\left[(n+m)\Psi_{00}\right]+l\ f_0,
	\label{eq:calcs_eigenspectra}
\end{equation}
for $l\in\mathcal{N}$. 
A cavity with an odd number of reflections per round trip splits the degeneracy between the modes with even 
and odd $n$ by adding an extra $\pi$ round trip phase for odd $n$ modes. 
In this case the expression must be modified slightly to
\begin{equation}
	 f_{nm}=\frac{c}{2\pi L}\text{mod}_{2\pi}\left[(n+m)\Psi_{00}
		+\text{mod}_2(n)\pi\right]+l\ f_0.
\end{equation}	

%==================================================================================================
% First Order Resonance as a Thermal State Sensor
%==================================================================================================
\section{The first order resonance as a thermal state sensor}
\label{sec:thermal}

Equation \eqref{eq:calcs_eigenspectra} shows that the eigenspectrum of a spherical 
optical cavity is fully determined by the round-trip Gouy phase and the length of the cavity.  
The round-trip Gouy phase is fully determined, in turn, by the cavity geometry as shown by 
Eqns. \eqref{eq:calcs_gouy_fp} and \eqref{eq:calcs_g_factor} or \eqref{eq:calcs_gouy_complex}.  
Hence, a measurement of the location of any one of the higher-order resonances is a suitable probe 
for a change in the radius of curvature of one of the mirrors of a spherical optical cavity.  

Winkler et. al.~\cite{winkler_heating_1991} derive an approximate relationship for the change in 
sagitta, $\delta s$, of an optic which is heated by optical absorption;
\begin{equation}
	\delta s=\frac{\alpha}{4\pi\kappa}P_a,
\end{equation}
where $\alpha$ is the coefficient of thermal expansion, $\kappa$ is the thermal conductivity, and 
$P_a$ is the absorbed power.  
The change in sagitta can be expressed instead as a shift in the radius of curvature of the optic, $\delta R$, as
\begin{equation}
	\frac{1}{\delta R}=\frac{\alpha}{2\pi\omega^2\kappa}P_a,
	\label{eq:thermal_winkler_deltar}
\end{equation}
where $\omega$ is the beam size.  
This relationship is specific to the case of an optic which is heated at the surface by a beam being reflected 
from it, but a similar relationship exists for optics heated in the bulk\cite{winkler_heating_1991}.  

Since the dominant change in resonant optical cavities heated by optical absorption is a shift in 
the radii of curvature of the mirrors, a measurement of the frequency of any one of the HOM 
resonances is sufficient to characterize the thermal state of the cavity.  
In the following sections we exploit this fact in two particular ways; we use it to 
measure the absorption of the aLIGO IMC, and we use it to generate an 
error signal for the thermal state of the aLIGO PRC.

%==================================================================================================
% Input Mode Cleaner
%==================================================================================================
\section{Input Mode Cleaner}
\label{sec:imc}

\begin{figure}[t]
	\centering
	\includegraphics[width=0.9\columnwidth]{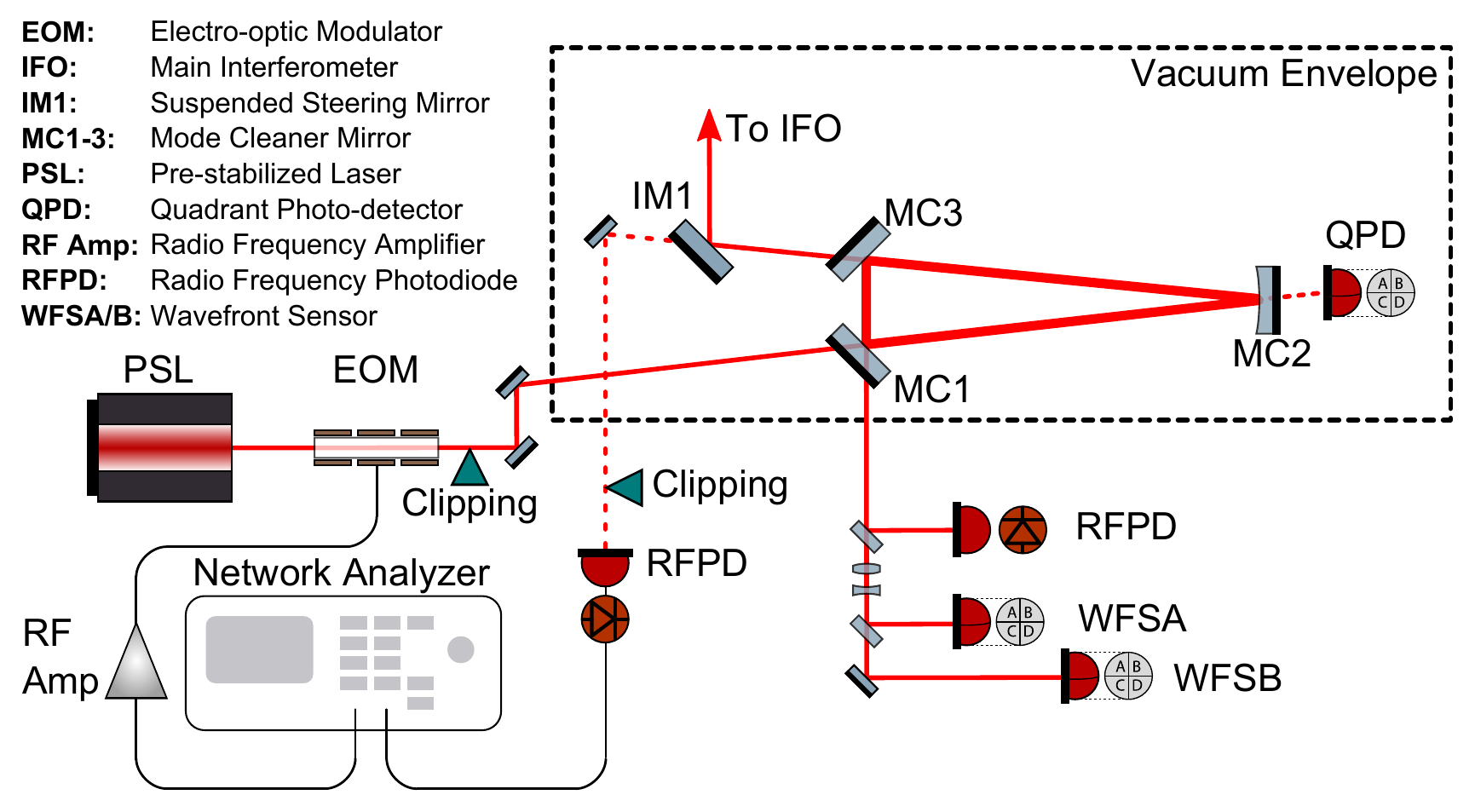}
	\caption{The experimental setup of the IMC eigenspectrum measurement.  
		The probe beam is generated by adding phase sidebands to the resonant carrier beam with the 
		aLIGO electro-optic modulator, and the signal is generated by demodulating in transmission.}
	\label{fig:imc_experimental_setup}
\end{figure}

The thermal characterization technique was first applied to the input mode cleaner 
(IMC) at the LIGO Livingston Observatory.  
The IMC is an in-vacuum, suspended, triangular optical cavity whose role in the aLIGO 
detectors is, amongst others, to suppress higher-order spatial modes on the input beam before 
injection into the main interferometer.  
Figure \ref{fig:imc_experimental_setup} shows the experimental layout.  
The IMC is formed by the three mirrors MC1, MC2, and MC3, and is outfitted with a resonant RF 
photodiode (RFPD) for length sensing and two differential wavefront sensors, WFSA and WFSB, for alignment 
sensing~\cite{anderson_alignment_1984}.  
An active control system uses these signals to maintain the IMC on resonance for the fundamental 
mode of the carrier beam as well as to keep it aligned to this mode.  
The length control loops have a unity gain frequency (UGF) of $70\,\text{kHz}$ while the angular 
loops have a UGF of $500\,\text{mHz}$.  

%^^^^^^^^^^^^^^^^^^^^^^^^^^^^^^^^^^^^^^^^^^^^^^^^^^^^^^^^^^^^^^^^^^^^^^^^^^^^^^^^^^^^^^^^^^^^^^^^^^
%^^^^^^^^^^^^^^^^^^^^^^^^^^^^^^^^^^^^^^^^^^^^^^^^^^^^^^^^^^^^^^^^^^^^^^^^^^^^^^^^^^^^^^^^^^^^^^^^^^

While the cavity is maintained on resonance by the control system, a network analyzer and RF 
amplifier are used to add a pair of RF sidebands to the carrier beam by phase modulation applied 
with an electro-optic modulator (EOM).  
Downstream of the EOM the beam is partially occulted by placing a small metal pin 
into the beam, generating HOMs on both the sideband and the carrier out of the 
initially mostly HG$_{00}$ beam.  
A sample of the light in transmission of the IMC is picked off and brought out of the vacuum 
system where it is occulted in a similar manner and focused onto a broadband RFPD.  
The signal from this RFPD is fed to the network analyzer where it is demodulated at the RF 
modulation frequency. 

%^^^^^^^^^^^^^^^^^^^^^^^^^^^^^^^^^^^^^^^^^^^^^^^^^^^^^^^^^^^^^^^^^^^^^^^^^^^^^^^^^^^^^^^^^^^^^^^^^^
\begin{figure}[t]
	\centering
	\includegraphics[width=0.8\columnwidth]{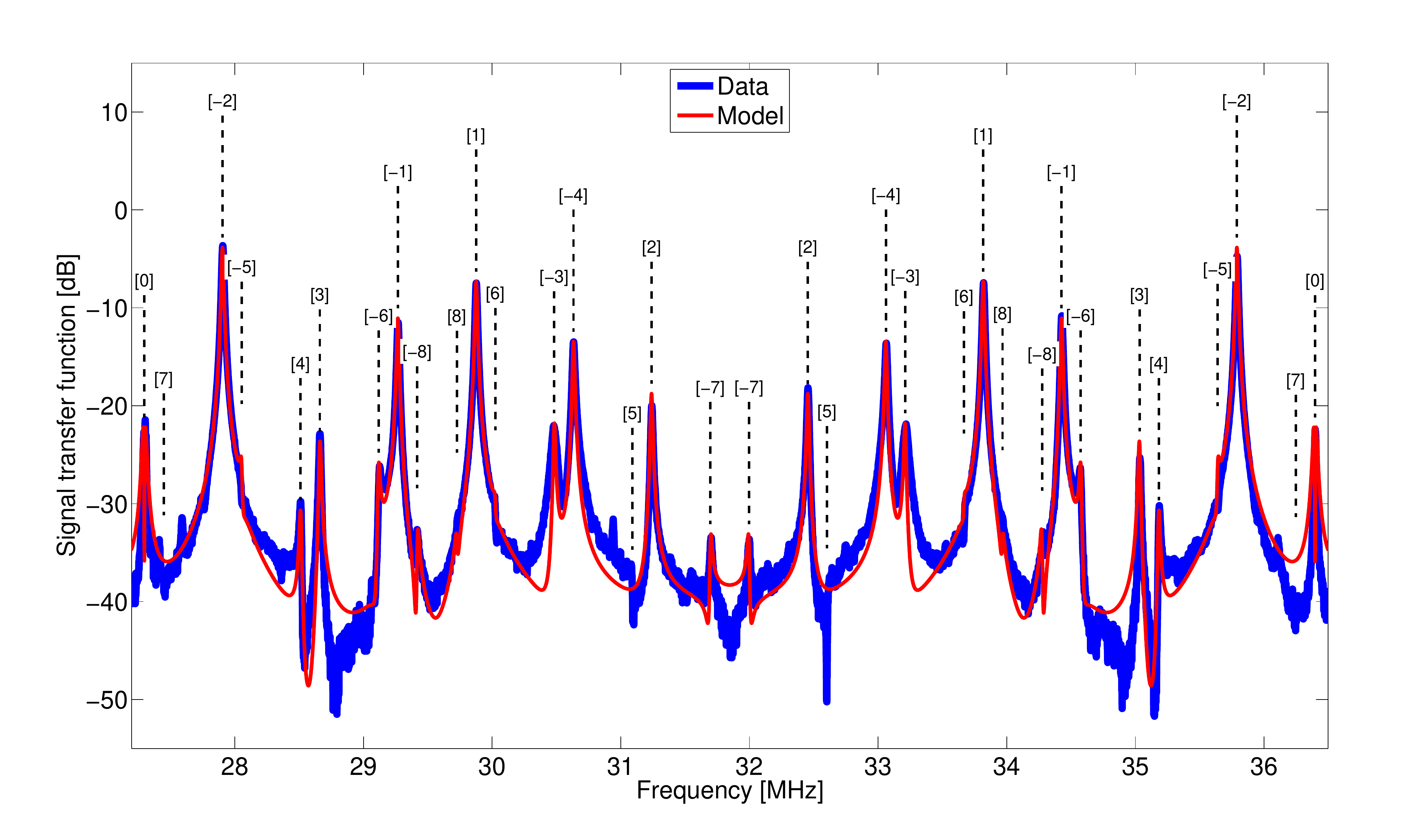}
	\caption{The magnitude of an example sweep across a full free spectral range of the input mode 
		cleaner (IMC) together with a \textsc{Finesse} model.  
		The peaks are labeled with their spatial mode orders ($n+m$) with a minus sign indicating the 
		split peaks of the odd orders ($n$).
		Since the IMC eigenspectra is probed with phase modulation, each peak appears twice from being 
		excited independently by the upper and lower sideband.}  
	\label{fig:imc_full_sweep}
\end{figure}
%^^^^^^^^^^^^^^^^^^^^^^^^^^^^^^^^^^^^^^^^^^^^^^^^^^^^^^^^^^^^^^^^^^^^^^^^^^^^^^^^^^^^^^^^^^^^^^^^^^		

Figure \ref{fig:imc_full_sweep} shows data taken with this measurement across a 
full FSR of the IMC.  
The data are shown together with a \textsc{Finesse}~\cite{freise_frequency-domain_2004} model of 
the measurement and labels for the particular spatial mode order of each of the peaks.  
This figure shows that modes up to order four are visible with this measurement without 
using a more complicated strategy for HOM creation and detection.  
This experiment was used to track the HG$_{10}$ resonance of the IMC while the input 
power was cycled. 
The shift in this peak was used in infer the absorption in the IMC.  

Although the Winkler approximation, discussed in Section \ref{sec:thermal}, is useful for 
illustrating the physics behind the shift in first-order mode resonance and thus the measurement 
principle, in practice the situation is more complicated. 
The thermal distortion produced by absorption of the beam on the coating is not a strictly parabolic 
deformation.
The effective curvature seen by each higher-order mode is different due to their different intensity 
profiles.  
As a general rule of thumb; the higher the mode order, the less effective curvature change it 
experiences for a thermal distortion produced by absorption of the fundamental mode.  
A numerical model was therefore used to convert shifts in the HG$_{10}$ resonance into absorption 
at the optic surfaces in order take such effects into account.

The numerical model begins by calculating the deformation of the surface of the mirrors using a 
finite element model in the commercial software package COMSOL.  
The surface deformation is calculated for various levels of absorbed power ranging from 1 mW to 
20 mW.  

These deformation maps are used together with the measured phase maps of the optics in an fast 
Fourier transform (FFT) 
based simulation which uses FFT beam propagation to calculate the stable mode in an optical cavity.  
Once the stable mode is calculated, the location of the first order resonance is identified by 
numerically adding RF sidebands to the beam similar to the real experiment.  
The numerical model assumes that the absorption is the same at all three mirrors.

Table \ref{tab:imc_absorption} shows the results of this measurement during various repetitions 
taken over the course of nearly a year and a half. 
The final column shows the absorption inferred from the numerical model 
which assumes that the absorption is the same at each mirror. Throughout the time period over 
which these measurements were made, the vacuum chambers in which the IMC optics are suspended 
were variously vented, and the optics cleaned before re-evacuating the chambers. The apparent changes 
in absorption over time may therefore be attributable to the different states of the optics at the time of 
each measurement. 

%^^ IMC: Losses: Absorption Measurement Results ^^^^^^^^^^^^^^^^^^^^^^^^^^^^^^^^^^^^^^^^^^^^^^^^^^^
\begin{table}[t]
	\centering
	\renewcommand{\arraystretch}{1.2}
	\begin{tabular}{lrrrr}
		\hline
		Date &  Power (W)  &  $f_{10}$ (Hz) & 	$\Delta f_{10}$ (Hz) & 
			  Abs. (ppm/mir.)  \\
		\hline
		\multirow{2}{*}{1/17/2013} & 	3.11 & $29,266,891\pm60$ & \multirow{2}{*}{$6,230\pm69$} 
			& \multirow{2}{*}{$2.39\pm0.02$}\\
		& 30.5 & $29,273,121\pm35$ & & \\
		\hline
		\multirow{2}{*}{7/23/2013} & 	0.517 & $29,268,352\pm65$ & \multirow{2}{*}{$245\pm85$} 
			& \multirow{2}{*}{$1.50\pm0.46$}\\
		& 1.03 & $29,268,597\pm56$ & & \\
		\hline
		\multirow{2}{*}{8/6/2013} & 	0.203 & $29,267,831\pm14$ & \multirow{2}{*}{$358\pm15$} 
			& \multirow{2}{*}{$1.42\pm0.06$}\\
		& 1.01 & $29,268,189\pm6$\hspace{4pt}\phantom{} & & 	\\
		\hline
		\multirow{2}{*}{9/30/2013} & 	1.07 & $29,266,056\pm39$ & \multirow{2}{*}{$1,175\pm54$} 
			& \multirow{2}{*}{$1.53\pm0.06$}\\
		& 3.08 & $29,267,230\pm37$ & & \\
		\hline
		\multirow{2}{*}{6/17/2014} & 1.79 & $29,269,212\pm23$ & \multirow{2}{*}{$1,585\pm42$}
			& \multirow{2}{*}{$0.50\pm0.01$}\\
		& 10.2 & $29,270,797\pm36$ & & \\
		\hline
	\end{tabular}
	\caption{The data and inferred absorption from numerous repetitions of the Gouy phase absorption 
		measurements in the LLO IMC. The Power and $f_{10}$ columns show the input power level and location 
		of the TEM$_{10}$ peak while the $\delta f_{10}$ and Abs. columns show the shift in this peak 
		between power levels and the inferred absorption.}
	\label{tab:imc_absorption}
\end{table}
%^^^^^^^^^^^^^^^^^^^^^^^^^^^^^^^^^^^^^^^^^^^^^^^^^^^^^^^^^^^^^^^^^^^^^^^^^^^^^^^^^^^^^^^^^^^^^^^^^^

%==================================================================================================
% Power Recycling Cavity
%==================================================================================================
\section{Power Recycling Cavity}
\label{sec:prc}

A similar technique was applied to the power recycling cavity (PRC) of the aLIGO interferometers 
in order to sense their thermal state. 
The power recycling technique, employed in both initial LIGO and aLIGO, 
redirects light reflected from the Michelson interferometer back into the interferometer 
with the advantageous effects of increasing the circulating power and filtering the noise of the 
input beam while maintaining the bandwidth of the instrument's response to gravitational wave 
signals~\cite{harry_advanced_2010}.  
The increase in power reduces the shot noise equivalent strain, thus increasing the detector's 
sensitivity.  

\begin{figure}[t]
	\centering
	\includegraphics[width=0.8\columnwidth]{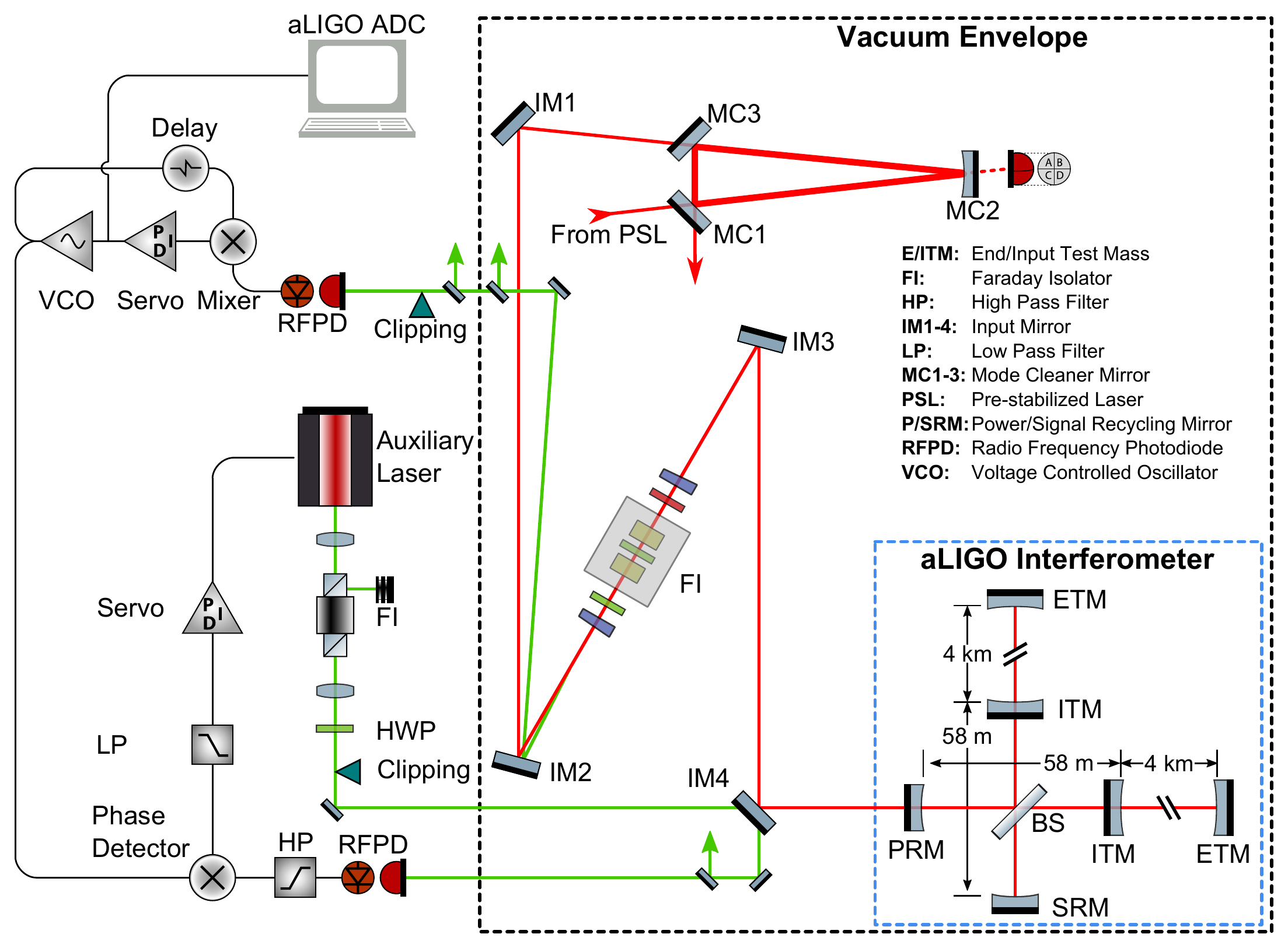}
	\caption{The experimental setup of the PRC thermal state sensor.}
	\label{fig:prc_experimental_setup}
\end{figure}

This increase in circulating power is not, however, without cost.  
Small yet finite absorption in the test masses of the interferometer coupled with non-zero 
thermo-elastic and thermo-optic coefficients cause deleterious thermal effects when operating the 
interferometer at high power.  
Both the input test masses (ITMs) and end test masses (ETMs) suffer from a shift in their radii 
of curvature as was described in Section \ref{sec:thermal}.  
In addition, the non-zero thermo-optic coefficient leads to a significant thermal lens in the test mass 
substrates. The substrate thermal lens in the ITMs plays a role in determining the geometry of the 
PRC because the ITM substrates are situated within the PRC 
formed between the power recycling mirror (PRM) and reflective surfaces of both ITMs (see Figure 
\ref{fig:prc_experimental_setup}). 
Both the thermo-optic and thermo-elastic effects act, to first order, to change the effective 
radii of curvature of the mirrors as sensed by the light circulating in the PRC. 
Hence, tracking of the first order resonance provides an excellent method of monitoring these 
thermal effects. 

Figure \ref{fig:prc_experimental_setup} shows the layout of the experiment used to track the first 
order resonance of the PRC.  
It was necessary to inject an auxiliary laser into the interferometer after the IMC, because 
the IMC strips off any sidebands added to the input beam that are not integer multiples of its FSR.  
This auxiliary laser is phase locked to the main interferometer beam with an offset which is set by the RF 
output of a voltage controlled oscillator (VCO) with a center frequency at 27.3\,MHz, a multiple of the 
cavity FSR.  
The phase locking is accomplished by combining a pickoff of the main beam and the auxiliary beam 
on an RFPD.  
The beatnote from this RFPD is high-pass filtered and mixed with the RF output of the VCO using a 
Minicircuits ZRPD-1 phase detector.  
The low-pass filtered output from this phase detector is fed through a simple servo to the frequency 
actuator in the auxiliary laser, thereby closing a phase-locked loop with a UGF 
of 50\,kHz.   

The auxiliary laser beam is occulted before being injected into the interferometer to generate HOMs, 
and the reflected beam from the interferometer, including both the carrier and sideband, is picked 
off by the in-vacuum Faraday isolator and a sample of the light is brought to an out-of-vacuum 
optical table.  
This sample of the reflected beam is occulted in a similar manner to the input beam before being 
focused onto a broadband RFPD.  
The signal from the RFPD is demodulated against the VCO signal which is slightly delayed in order 
to ensure that the output of the mixer is in the optimal quadrature for sensing the beat signal between 
the fundamental and first order modes.
This signal is fed through a simple $f^{-1}$ servo whose output drives 
the VCO, closing a feedback loop with a UGF of 1\,Hz.  
The output of the servo is proportional to the frequency output of the VCO and is recorded into 
the aLIGO digital controls system.  

%^^^^^^^^^^^^^^^^^^^^^^^^^^^^^^^^^^^^^^^^^^^^^^^^^^^^^^^^^^^^^^^^^^^^^^^^^^^^^^^^^^^^^^^^^^^^^^^^^^
%^^^^^^^^^^^^^^^^^^^^^^^^^^^^^^^^^^^^^^^^^^^^^^^^^^^^^^^^^^^^^^^^^^^^^^^^^^^^^^^^^^^^^^^^^^^^^^^^^^			

The result of this experimental setup is that the frequency of the auxiliary laser is locked to 
the first order resonance of the aLIGO PRC. 
The frequency of this resonance relative to the fundamental resonance is a tracer for the thermal 
state of the interferometer, and can be used as a feedback signal to the aLIGO thermal compensation 
systems (TCS) to maintain the interferometer in a particular thermal state. 
Using the Winkler approximation described in Section \ref{sec:thermal} together with the ray 
matrix techniques described in Section \ref{sec:calcs} and assuming that the 
effective radius of curvature of the two ITMs shift in unison, we can derive the round-trip Gouy phase 
change of the first order mode to be
\begin{equation}
	\Delta\Psi_{00}=0.701\ \text{rad}+0.014\ \frac{\text{rad}}{\mu\text{D}}\ \delta p,
	\label{eq:prc_curvature_shift}
\end{equation}	
where $\delta p=\frac{1}{\delta R}$ is the change in the mirror curvature.  
This can be converted to the frequency shift of the first order mode using 
Eqn.~\eqref{eq:calcs_eigenspectra} where the round-trip length of the PRC is $L=115.232\ \text{m}$.

%^^^^^^^^^^^^^^^^^^^^^^^^^^^^^^^^^^^^^^^^^^^^^^^^^^^^^^^^^^^^^^^^^^^^^^^^^^^^^^^^^^^^^^^^^^^^^^^^^^
\begin{figure}[t]
	\centering
	\includegraphics[width=0.8\textwidth]{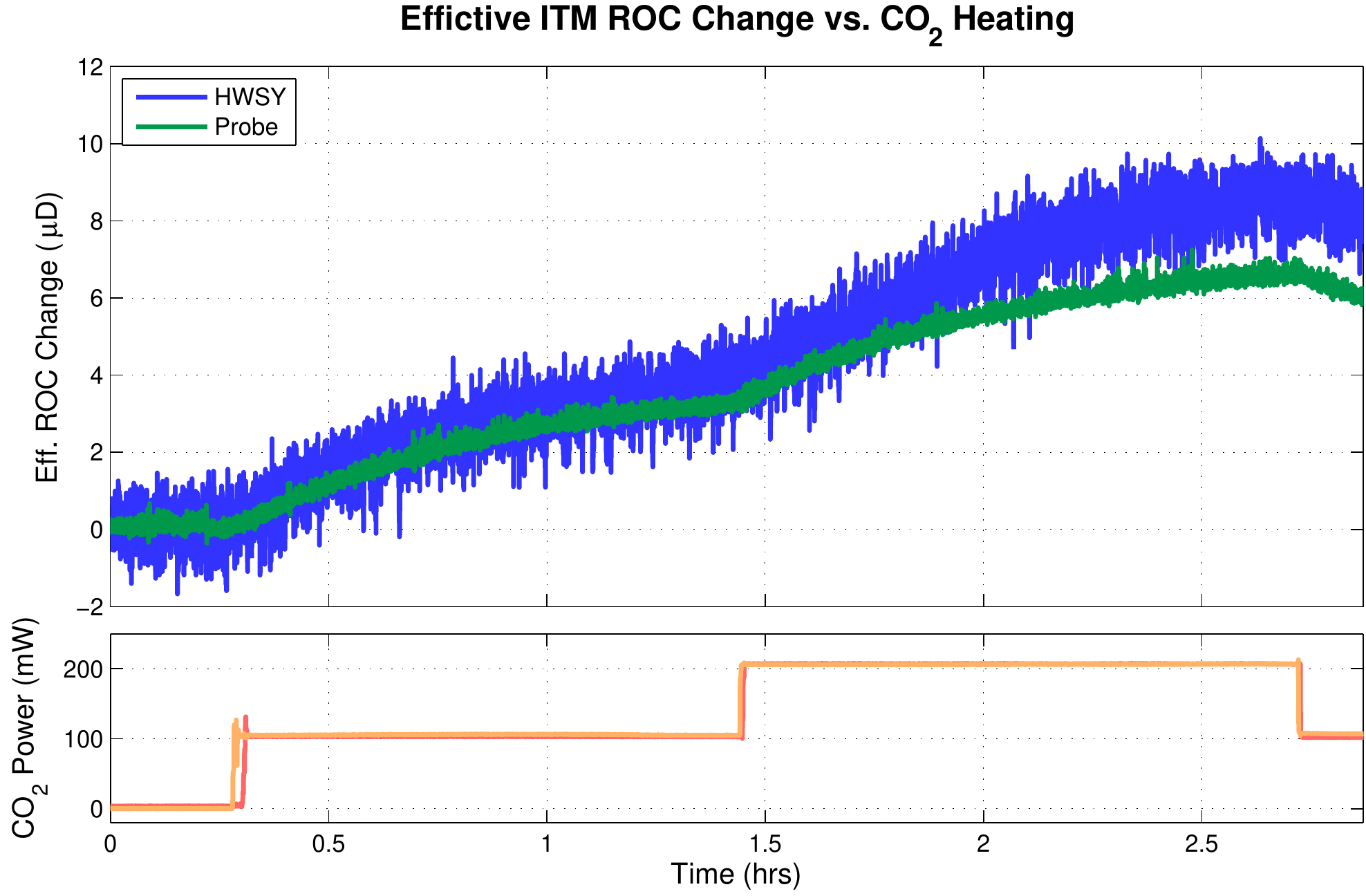}
	\caption{The shift in the resonant frequency of the first order excited mode is shown while 
		the cavity is heated with CO$_2$ lasers projected onto the test masses.  
		The shift in the resonant frequency is converted into a shift of the radius of curvature of 
		the test mass through a ray matrix model described in the text.  
		Also shown is the effective ROC shift in the test masses as measured by a Hartmann wavefront 
		sensor.}
	\label{fig:prc_thermal_tracking}
\end{figure}
%^^^^^^^^^^^^^^^^^^^^^^^^^^^^^^^^^^^^^^^^^^^^^^^^^^^^^^^^^^^^^^^^^^^^^^^^^^^^^^^^^^^^^^^^^^^^^^^^^^

Figure \ref{fig:prc_thermal_tracking} shows the result of an experiment in which optical 
absorption in the PRC is simulated by applying equal amounts of heat to the two transmissive 
compensation plates which are situated directly in front of the input
test masses with two CO$_2$ lasers operating at a wavelength of 10.6~$\mu$m.  
At this wavelength nearly all of the optical power applied by the lasers is absorbed at the compensation plates.  
The size of the beam projected onto the compensation plates is roughly matched to the size of the science beam 
and thus provides an excellent stand-in for optical absorption. The compensation plates are located extremely close 
to the ITMs relative to the beam Rayleigh range and thus lensing in these plates causes 
an almost identical change in the PRC eigenmode to lensing in the ITMs themselves. Henceforth we therefore refer to 
lensing produced in the compensation plates simply as lensing in the ITM.

During the experiment, the change in the radius of curvature (ROC) of the ITMs as measured 
by the Advanced LIGO Hartmann wavefront sensors~\cite{brooks_ultra-sensitive_2007} (HWS) 
was also recorded.  
These devices use an aperture array in front of a CCD to image the wavefront of a beam reflected 
from each ITM. 
The optical setup of the HWS is designed such that the imaged wavefront is in an image plane of 
the ITM so that a change in the effective ROC of the ITM shows up as a change in the wavefront 
curvature at the HWS.  

Equation \ref{eq:prc_curvature_shift} was used to convert the frequency shift of the first order 
resonance into the ROC shift of the ITMs.  
Figure \ref{fig:prc_thermal_tracking} shows that the ROC shift measured by the first order 
resonance and that measured by the HWS are in good agreement.  
The residual disagreement is likely attributable to errors in the HWS calibration.  
This highlights an advantage of the first order resonance tracking over that of the HWS.  
Tracking of the first order resonance measures precisely the effective radius of curvature which 
is seen by the main resonant beam.  
The significantly better SNR of the first order resonance over that of the HWS showcases another of 
its advantages. One disadvantage however is that the first order mode tracking method requires the 
PRC to be locked, whereas the HWS can operate independent of the resonance condition of the interferometer. 

%==================================================================================================
% Conclusion and Outlook
%==================================================================================================
\section{Conclusion and Outlook}
\label{sec:conclusion}

In conclusion, tracking of the HG$_{10/01}$  mode resonance was shown to provide an excellent tracer for the 
thermal state of an optical resonator.  
This method was applied to measure the optical absorption of the input mode cleaner (IMC) at the LIGO 
Livingston Observatory.  
Although ultra-low absorption in the IMC is not essential to the overall operation of gravitational 
wave interferometers, monitoring the absorption over time provides an excellent method of tracking 
the degradation, or lack thereof, of the optical coatings over the long time periods for which 
these devices are designed to operate.  
It can also be used as a method of validating the cleanliness of the vacuum system after 
incursions for repairs or maintenance.  

It was also shown that real-time tracking of the thermal state of the resonant cavities of interferometric  
gravitational wave detectors is possible by tracking the first-order resonance. 
This represents a significant breakthrough in thermal state sensing since all previous monitors 
have only been able to monitor the thermal state of individual mirrors.  
In contrast, tracking of the first excited mode provides a thermal error signal which is directly 
tied to the configuration of the interferometer.  

The technique applied to track the thermal state of the Advanced LIGO power recycling cavity (PRC) 
could be extended to sense the thermal state of the other resonant cavities in interferometric gravitational wave 
detectors.  
Doing so would provide independent thermal error signals for all of the important thermal degrees 
of freedom of these devices, thus making the process of compensating for thermal distortions more deterministic 
and repeatable than in previous states of operation.

%==================================================================================================
% Bibliography
%==================================================================================================
\bibliography{eigbib}

\end{document}